\documentclass[apj]{emulateapj}

\newcommand{\kpc}{{\rm kpc}}

\newcommand{\hmpc}{\ifmmode{h^{-1}\,\hbox{Mpc}}\else{$h^{-1}$\thinspace Mpc}\fi}
\newcommand{\etal}{et~al.}
\newcommand{\kms}{\ifmmode{\,\hbox{km}\,s^{-1}}\else {\rm\,km\,s$^{-1}$}\fi}
\newcommand{\msun}{{\rm\,M_\odot}}

\begin{document}
\title{Dwarf Galaxy Clustering and Missing Satellites}
\shorttitle{Dwarf Galaxy Clustering}
\shortauthors{Carlberg, Sullivan \& Le Borgne}
\author{
R.~G.~Carlberg\altaffilmark{1},
M.~Sullivan\altaffilmark{2}, and
D.~Le~Borgne\altaffilmark{3}, 
}

\altaffiltext{1}{Department of Astronomy and Astrophysics, University
  of Toronto, Toronto, ON M5S 3H4, Canada}
\altaffiltext{2}{University of Oxford Astrophysics, Denys Wilkinson
  Building, Keble Road, Oxford OX1 3RH, UK}
\altaffiltext{3}{Institut d'Astrophysique de Paris, UMR7095 CNRS, UPMC, 98bis boulevard Arago, F-75014 Paris, France}

\email{carlberg@astro.utoronto.ca }

\begin{abstract}
At redshifts around 0.1 the CFHT Legacy Survey Deep fields contain some $6\times 10^4$ galaxies spanning the mass range from $10^5$ to $10^{12}\msun$.  We measure the stellar mass dependence of the two point correlation using angular measurements to largely bypass the errors, approximately 0.02 in the median, of the photometric redshifts. Inverting the power-law fits with Limber's equation we find that the auto-correlation length increases from a very low $0.4\hmpc$ at $10^{5.5}\msun$  to the conventional $4.5\hmpc$ at $10^{10.5} \msun$.  The power law fit to the correlation function has a slope which increases from $\gamma\simeq 1.6$ at high mass to $\gamma\simeq 2.3$ at low mass. The spatial cross-correlation of dwarf galaxies with more massive galaxies shows fairly similar trends, with a steeper radial dependence at low mass than predicted in numerical simulations of sub-halos within galaxy halos. To examine the issue of "missing satellites" we combine the cross-correlation measurements with our estimates of the low mass galaxy number density. We find on the average there are $60\pm20$ dwarfs in sub-halos with $M({\rm total}) > 10^7\msun$ for a typical Local Group $M(\rm total)/M(\rm stars)=30$, corresponding to $M/L_V \simeq 100$ for a galaxy with no recent star formation. The number of dwarfs per galaxy is about a factor of two larger than currently found for the Milky Way.  Nevertheless, the average dwarf counts are about a factor of 30 below LCDM simulation results. The divergence from LCDM predictions is one of slope of the relation, approximately $dN/d\ln{M}\simeq -0.5$ rather than the predicted $-0.9$, not sudden onset at some characteristic scale. The dwarf galaxy star formation rates span the range from passive to bursting, which suggests that there are few completely dark halos. 
\end{abstract}


\section{INTRODUCTION}
\nobreak
The detailed physics of whatever constitutes the Cold Dark Matter (CDM) may manifest itself in the sub-megaparsec regime of the LCDM density perturbation spectrum from which low mass dark matter halos originate. Low stellar mass galaxies are expected to inhabit proportionally low mass dark halos. The available evidence supports this view, although dwarf galaxies generally have a higher total mass to stellar light ratio than higher mass galaxies (reviewed in Mateo 1998).  Clustering theory predicts that the two-point correlation of dark matter halos relative to the entire density field, that is, bias, decreases for lower masses to become slightly less clustered than the mean mass distribution \citep{K:84,ST:99}. The easiest clustering measurement to make, since it has the largest signal, is to probe nonlinear clustering where the prediction is not as clear, since it requires numerical simulations on scales which are not currently well resolved for representatively large cosmological volumes. A specialized case of nonlinear clustering is the distribution of dwarf galaxies within the halos of larger galaxies, such as the Milky-Way, which is well resolved in n-body simulations.  As a consequence of the high central density of low mass halos, their central regions are very resistant to dissolution when they merge into high mass halos, although their outer regions are tidally stripped away. The outcome is a robust prediction that there should be hundreds of low mass sub-halos within the halo of large galaxy, which is at least a factor of ten higher than the visible dwarf galaxies \citep{Moore:99, Klypin:99, SG:07} of the Milky-Way.

The problem of relating light to mass for dwarf galaxies clouds the comparison to dark matter only simulations. The first issue is that dwarf galaxies have been studied in detail for a single galaxy, the Milky Way. The current census finds that there are about two dozen known Milky Way dwarf galaxies, with many of them having been discovered only in the last decade. Under the assumption of a uniform sky distribution (up to a factor of five correction) the total numbers are estimated to be about 30 to 50 to $M_V$ of about -8 and -4, respectively \citep{Mateo:98,SG:07}. 

Measurements of velocities for hundreds of individual stars within a goodly number of Milky-Way dwarfs allow an accurate estimate of the mass contained within the orbits of the stars for those galaxies. Unfortunately the mass outside the distribution of the stars is essentially unconstrained, even if one makes specific assumptions of the properties of the dark halo \citep{PMN}. One surprising outcome is that the mass enclosed within the stellar distribution is usually close to $10^7\msun$ \citep{Mateo:98, Strigari08} over a wide range of dwarf galaxy luminosities. The kinematic modelling finds that the central density distribution appears to have a core with a universal maximum value for the central density \citep{Gilmore:07}, whereas CDM halos should have cusped central density distributions. A dynamical upper limit to the total bound mass is that if it becomes too large then their gravitational heating of the thin disk of the Milky Way would become unacceptably large. Some authors have viewed substructure as a desirable source of disk heating \citep{Benson:04}. In general, the limit on  disk requires that if the entire predicted satellite population is present, then its radial distribution must be less centrally concentrated than the overall mass distribution \citep{Font:01, Ardi:03}.  In spite of these uncertainties the conclusion remains that the Milky Way dwarf galaxies at a total mass of $10^7\msun$ are a factor of ten less numerous than the predicted numbers of sub-halos \citep{Strigari07}.

Here we use the CFHT Legacy Survey deep imaging survey to study a large sample of field dwarfs to complement the Local Group studies. Only photometric information is available for each of these small, faint galaxies. However, we show that the photometric redshifts turn out to be sufficiently accurate that we can reliably place these objects into wide redshift bins to measure their volume density and their two-point correlation function. We show that we can recover standard clustering estimates for massive galaxies to validate our measurement techniques. Then we turn to the cross-correlation of low and high mass galaxies. To examine the missing satellite problem we combine the radial cross-correlation functions with our estimate of the volume density of low mass galaxies to derive the numbers of satellites. The measurements in this paper use a Hubble constant of $100h \kms$~Mpc$^{-1}$ in a flat $\Omega_M=0.3$ universe. For comparison to a specific simulation we use $h=0.73$ and $\Omega_M=0.24$, although these parameter changes do not make much of a difference at the low redshifts of this paper.

\section{DATA and METHODS}
\label{sec:sample}
\nobreak 
The deep fields of the CFHT Legacy Survey provide [$u_M,g_M,r_M,i_M,z_M$] images to a depth of $i_M=$ 26.5 mag in the Megaprime filter system (similar to the SDSS filters). For this study we use the D1, D2 and D4 fields, simply because we wanted an odd number of fields to allow median estimators to be used. The stacking combines better seeing exposures leading to stacks having images of 0.75\arcsec\ in $r_M, i_M, z_M$, 0.84\arcsec\ in $g_M$ and 0.93\arcsec\ in $u_M$. The $i_M$ filter is the deepest, with the $u_M$ and $z_M$ images having about 2/3 the number of objects that the $i_M$ catalogue has. The galaxy counts between $i_M=26$ and $26.5$ mag are rising approximately as $\log{N}=0.16 i_M$, slightly shallower than the $0.21$ seen in the Hubble Deep Field \citep{Williams:96}, indicating that incompleteness is just setting in. 

The PEGASE/ZPEG spectral energy distribution modelling program is used to derive indicative star formation rates (SFRs), stellar masses and photometric redshifts for all of the galaxies from the photometry \citep{LeBorgne:02, LeBorgne:04, Sullivan:06}. Our modelling uses the Rana \& Basu (1992) initial mass function which gave slightly better photometric redshifts than alternative IMFs. The particular modelling parameters adopted limits the maximum specific SFR to about $10^{-8} {\rm yr}^{-1}$, although this has no effect on our results. The modelled stellar masses depend on the IMF chosen with alternative IMFs giving systematic mass differences of about a factor of two. For instance, the Baldry \& Glazebrook (2003) IMF in ZPEG gives median galaxy masses that are on the average 40\% smaller than our adopted IMF. All of our clustering and density analysis uses mass bins a full decade wide to give good statistics and minimize the effect of the modelling uncertainties in shifting galaxies from one bin to another.

The $u_M$ calibration has been updated from the one in Sullivan et al (2006) which helps improve the photometric redshift accuracy. Comparison of the photometric redshifts to spectroscopic redshifts demonstrated a redshift accuracy of a median error of only 0.02, or 0.012/(1+z) \citep{Sullivan:06}. The redshifts are not sensitive to the IMF choice.
Redshifts derived from supernovae in some of the host galaxies provide redshifts for galaxies as faint as $i_M\simeq 27$ mag.  There are no spectroscopic redshifts for the faintest galaxies themselves, however the errors should be comparable to the brighter galaxies, although almost certainly somewhat larger due to increased photometric noise with possible systematic errors as a result of the lower metallicities of dwarf galaxies. To minimize the impact of redshift errors on our clustering measurements we make angular clustering measurements over redshift bins significantly wider than our redshift error. Comparison of the redshifts from different parameter choices for ZPEG finds that at least 80\% of the redshifts, and often more than 90\%, will be within 0.02 of the adopted redshift, with the other distributed in long, low amplitude tails. 

\begin{figure}
\epsscale{0.90}
\plotone{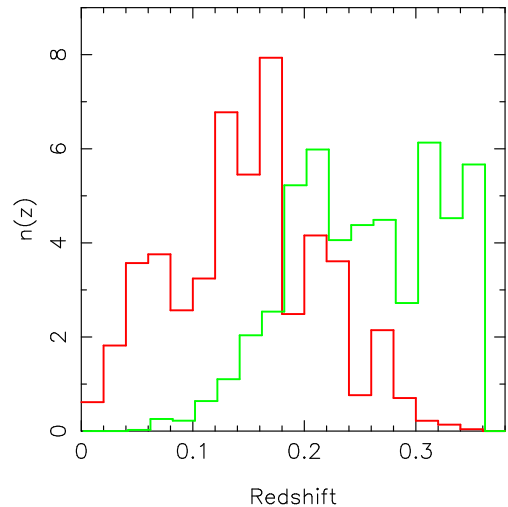}
\caption{The redshift distribution of the $10^5-10^6 \msun$ (red) and $10^{10}-10^{12}\msun$ galaxies. The distributions are normalized $\sum n(z)\Delta z=1$. 
\label{fig_nz}}
\end{figure}

\begin{figure}
\epsscale{0.90}
\plotone{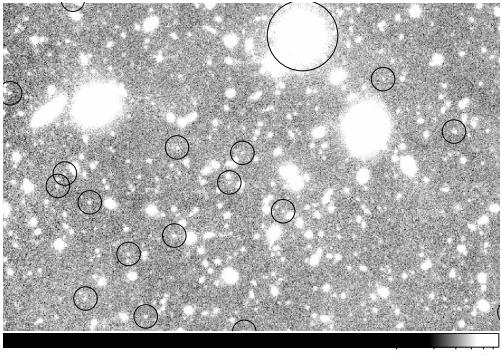}
\caption{A section of the D1 field with 3\arcsec\ circles around $10^5-10^6 \msun$ and 9\arcsec\ circles around galaxies more massive than $10^9\msun$ in the $z=0.1-0.18$ range. There are about 25 seeing disks of 0.75\arcsec\ radius per detected galaxy (at any redshift) in this field.
\label{fig_ds9}}
\end{figure}

The magnitude limit of $i_{AB}=26.5$ at redshift 0.15 corresponds to galaxies with absolute magnitudes of $M_i \le  -12$. Figure~\ref{fig_nz} shows the redshift distribution of galaxies in the range $10^5-10^6\msun$ and the rising redshift distribution of the more massive galaxies $10^{10}-10^{12}\msun$. The low mass galaxy numbers initially rise as would be expected for a complete sample, however they begin to fall below a uniform density rise of $z^2\Delta z$ beyond redshift 0.05 as galaxies with low star formation rates, hence lower luminosity at fixed mass, fall out of the sample. For a cross-correlation of high and low mass galaxies we need to select a redshift range which balances the two populations. Given the drop in low mass galaxy numbers beyond $z=0.18$ we use that as our upper redshift. A small section of the sky in the D1 field is shown in Figure~\ref{fig_ds9} with galaxies in the $z=0.1-0.18$ range with masses $10^5-10^6\msun$ and $10^9-10^{12}\msun$ marked. In total there are approximately $5\times10^5$ detectable objects in this field of 0.85 square degrees after masking, putting the low mass galaxies at the confusion limit \citep{Scheuer:57, Hogg:01}. 

\begin{figure}
\epsscale{0.90}
\plotone{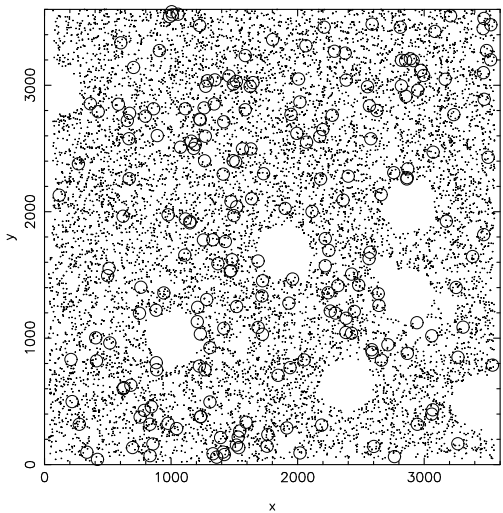}
\caption{The sky locations in field D1 of the $10^{5-6}\msun$ and $10^{10-12}\msun$ galaxies selected to be in the redshift range 0.1 to 0.18. 
\label{fig_D1}}
\end{figure}

The relative positions in the entire D1 field of galaxies in the mass ranges $10^5-10^6\msun$ and $10^{10}-10^{12}\msun$ in the selected redshift range are shown in Figure~\ref{fig_D1}. At redshift 0.15 the field has a transverse width of about 10\hmpc. The holes mask out the bright stars along with bogus objects and varying background light near them. The clustering pattern of the high mass galaxies is clearly evident. 

We will measure the clustering with the angular two-point cross-correlation function. The two-point function gives the sky density of galaxies of type 2 around those of type 1 at an angular separation $\theta$ as $n_{12}(\theta) = n_2[1+w_{12}(\theta)]$ (the expression is symmetric in the types). We use the Landy-Szalay (1993) estimator as modified to have the required symmetry in the two populations \citep{Blake:06}. For two samples of galaxies, that share a common distribution area, 
\begin{equation}
w(\theta,\theta+\Delta\theta) = {{D_1D_2 - D_1R - D_2R+RR}\over{RR}},
\label{eq:LS}
\end{equation}
where $D_1D_2$ is the number of pairs between samples one and two that have angular separations between $\theta$ and $\theta+\Delta\theta$. For the auto-correlation the pairs are measured within the single sample of interest. The random sample, R, is a uniformly distributed sample of about $10^5$ points, generated with the same boundaries and masked regions as the galaxy samples, which in our case have the same underlying smooth distribution. The $D_1R$ and $D_2R$ terms are the pair counts with the random points and, $RR$, the random points with themselves. For the relatively weak clustering visible being measured the classical estimator, $DD/DR-1$, is substantially less accurate. Clustering measurements do not depend on sample completeness to be accurate, provided that any incompleteness is not selective in location.  

\section{CORRELATION RESULTS}

\begin{figure}
\epsscale{0.90}
\plotone{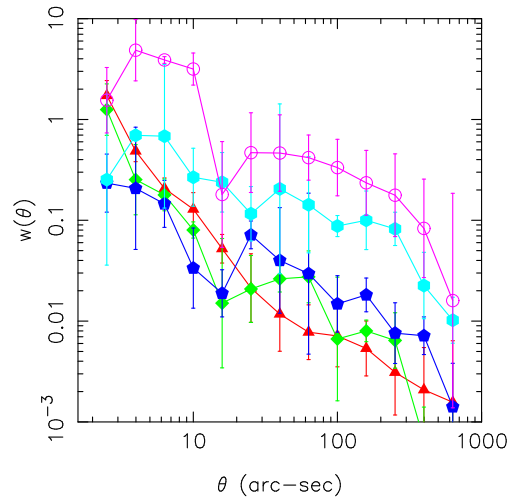}
\caption{The angular auto-correlation as a function of mass in bins of $10^5-10^6$, $10^6-10^7$, $10^7-10^8$, $10^8-10^9$ and $10^9-10^{12}\msun$ (triangles, diamonds, pentagons, hexagons and open circles, or, red, green, blue, cyan and magenta,  respectively). The errors of the mean are displayed.
\label{fig_w}}
\end{figure}

Figure~\ref{fig_w} shows the angular auto-correlation of the galaxies as a function of stellar mass. We display the median (the means are not significant different) of the D1, D2 and D4 field two-point function. The error bars are the errors in the means of the three fields. The pair counts in a single field for the smallest count bins are $10^2-10^3$, depending on mass, so the shot noise is usually under 10\%. The volume of each field is only about $10^4 h^{-3}$ Mpc$^3$ and the field has a linear span of about 6\hmpc. The fractional amplitude of the linear CDM density perturbation spectrum is such that a sphere of 8\hmpc\ radius has a fractional variance of 0.7-0.9, and the nonlinear variance will be even larger.  Accordingly we expect the field-to-field variance in excess of unity to overwhelm the relatively small shot noise in our measurement. 

\begin{figure}
\epsscale{0.90}
\plotone{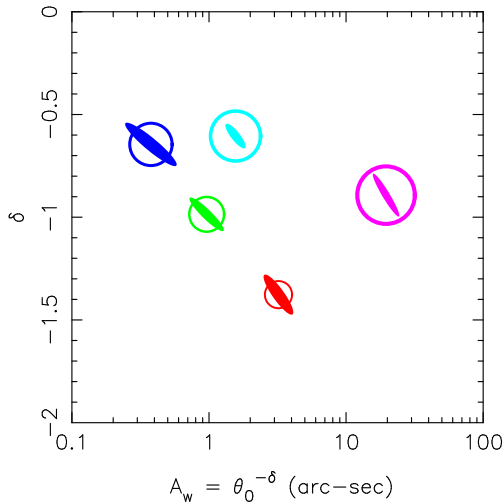}
\caption{The power law fits to the angular auto-correlation function. The circle size increases with the mass range of the galaxies for mass bins of $10^5-10^6$ (red), $10^6-10^7$(green), $10^7-10^8$ (blue), $10^8-10^9$ (cyan) and $10^9-10^{12}\msun$ (magenta) for the filled contours. The sizes of the circles centered on each point increases with the mass of the objects.
\label{fig_ga}}
\end{figure}

\begin{figure}
\epsscale{0.90}
\plotone{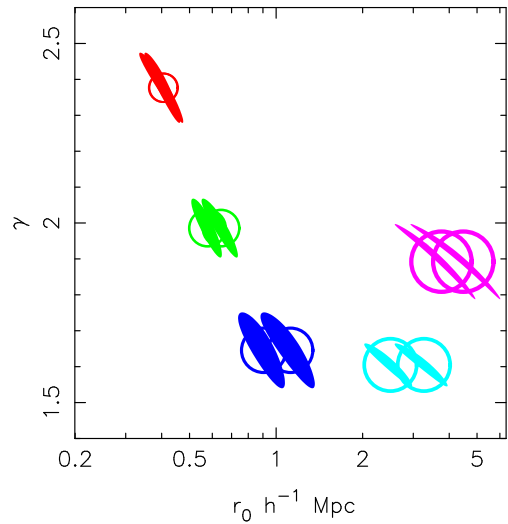}
\caption{The co-moving correlation length $r_0$ and the 3D power law, $\gamma$, derived using Limber's equation. The photometric redshift distribution is convolved with Gaussian redshift errors of 0.05 and 0.10 (believed to be an over-estimate) in redshift, with the larger error always giving a larger $r_0$. The $z=0.1-0.18$ sample is shown as filled contours, the $z=0.1-0.35$ with an open contour and the two bins of $z=0-0.05$ hatched. The mass bins are $10^5-10^6$ (red), $10^6-10^7$ (green), $10^7-10^8$ (blue), $10^8-10^9$ (cyan) and $10^9-10^{12}\msun$ (magenta). The circle size increases with the mass range of the galaxies for mass bins of $10^5-10^6$, $10^6-10^7$, $10^7-10^8$, $10^8-10^9$ and $10^9-10^{12}\msun$. 
\label{fig_rg}}
\end{figure}

We fit the power law, $w(\theta) = A_w \theta^\delta$, to the angular clustering measurements. The resulting fits, over the angular range 3-400\arcsec, along with their linear error contours are shown in Figure~\ref{fig_ga}. The point to point errors are sufficiently large that the power law is an acceptable fit to the data. Limber's equation inverts the angular clustering into a 3D clustering, $\xi(r)=(r_0/r)^\gamma$, with $A_w = \int n^2(z)r_0^\gamma r(z)^{1-\gamma} c H(z) \, dz/[\int n(z)\, dz]^2$, where we assume that the clustering is fixed in co-moving co-ordinates, $r(z)$. We take the true $n(z)$ to be a convolution of the photometric $n(z)$ with a Gaussian redshift distribution having a $\sigma_z$ of 0.05 or 0.10, mainly to demonstrate that our results are not sensitive to redshift errors. Results for both $\sigma_z$ are shown in Figure~\ref{fig_rg}. We adopt $\sigma_z=0.05$ for our quoted results. The error matrix is calculated as the linear derivatives of the $[A_w,\delta]$ errors into $[r_0,\gamma]$, which in the $\log{r_0}$ axis of Fig.~\ref{fig_rg} become slightly curved ellipses. For large masses, $M \ge 10^9 \msun$, the auto-correlation length $r_0$ and power law index, $\gamma$, measured here are in good agreement with the values measured in field galaxy redshift surveys \citep{Maddox:90, Norberg:02, Zehavi:05}, giving us confidence in that the photometric redshifts are acceptably accurate.  

The low mass galaxy correlation length is impressively low, approximately $0.4\hmpc$ with $\gamma\simeq 2$ at a median redshift of 0.14. 
If we take the dark matter correlation to have an equivalent length of $5\hmpc$ with the same $\gamma$ then the equivalent bias value, $b(M)$, is 0.1 although it is important to recognize that the measurement applies only to the non-linear regime. Current simulations do not reach such low masses in a field volume. The extrapolation of the Sheth \& Tormen (1999) bias equation (their Eq.~12) to low mass at low redshifts corresponds to $\nu$ approaching zero, or, $b\simeq 1+(2p-1)/\delta-1$, where $p=0.3$ and $\delta_1\simeq 1.68$, giving $b\simeq 0.76$, far above what we measure for our low mass galaxies. The Sheth \& Tormen result was derived for higher masses and for large scale clustering of dark halos alone, and are not directly testable with our measurements. Hamana et al. (2001) address the radial dependence of bias and do find that the nonlinear bias for $10^{12}\msun$ halos on a scale of 1\hmpc\ does drop into the range of about 0.1 that we measure. A clear test awaits better resolved simulations.

The particular problem of the clustering of low mass galaxies in the halo of the Milky Way has been examined in considerable depth as a possible discrepancy of LCDM theory \citep{Klypin:99, Moore:99, VL1, Aquarius}. The simulations find hundreds of sub-halos orbiting within a large galaxy size halo which is a factor of at 3 to 10 more than the number of dwarf galaxies. Our sample allows us to investigate the missing satellite problem beyond the Milky Way. The first step is to measure the cross-correlation between low and high mass galaxies.

\begin{figure}
\epsscale{0.90}
\plotone{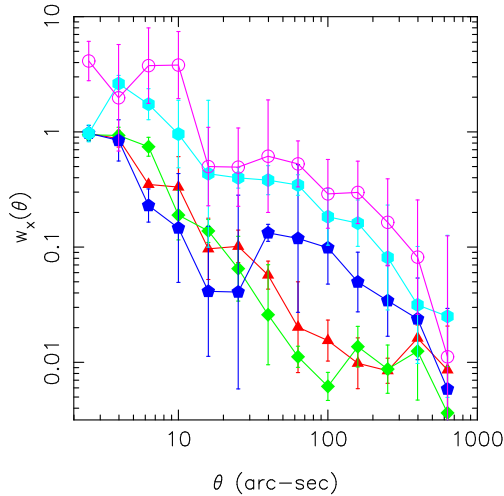}
\caption{The angular cross-correlation function of $10^{10}-10^{12}\msun$ galaxies with galaxies in the $10^5-10^6$ (red), $10^6-10^7$ (green), $10^7-10^8$ (blue), $10^8-10^9\msun$ (cyan) and $10^9-10^{10}\msun$ (magenta) range.
\label{fig_xcor}}
\end{figure}

\begin{figure}
\epsscale{0.90}
\plotone{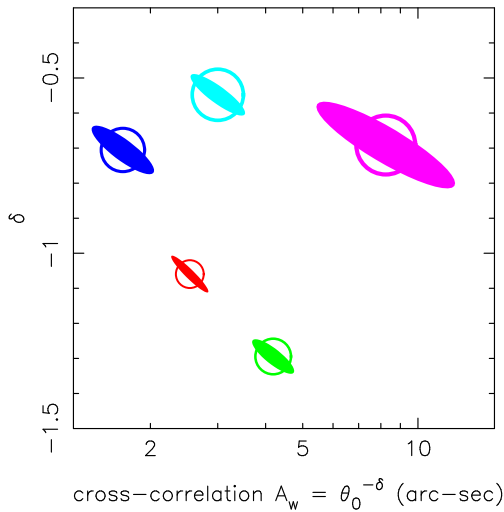}
\caption{The power law fits to the angular cross-correlation function of $10^{10}-10^{12}\msun$ galaxies with galaxies in the $10^5-10^6$ (red), $10^6-10^7$ (green), $10^7-10^8$ (blue), $10^8-10^9\msun$ (cyan) and $10^9-10^{10}\msun$ (magenta) range.
\label{fig_xga}}
\end{figure}

\begin{figure}
\epsscale{0.90}
\plotone{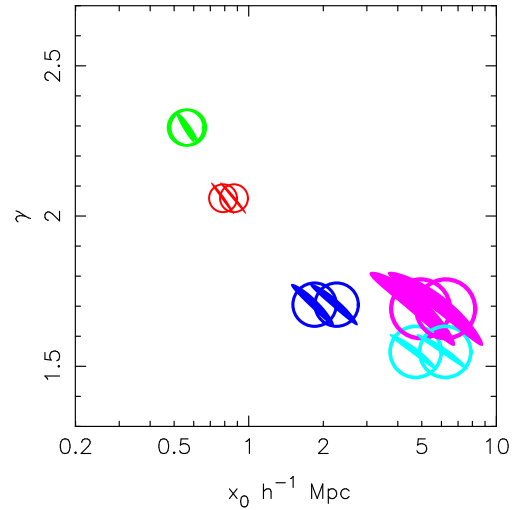}
\caption{The Limber's equation inversion of the angular cross-correlation to give the 3D cross-correlation length of $10^{10}-10^{12}\msun$ galaxies with galaxies in the $10^5-10^6$ (red), $10^6-10^7$ (green), $10^7-10^8$ (blue), $10^8-10^9\msun$ (cyan) and $10^9-10^{10}\msun$ (magenta) range.
\label{fig_xg}}
\end{figure}

The cross-correlation between two galaxy samples is measured with the same methods as we used for the auto-correlation functions. Figure~\ref{fig_xcor} shows the cross-correlations of low and high mass galaxies for the D1, D2 and D4 fields. Again the errors are dominated by the field to field variance, with the effective volume further reduced by the overlap integral of the two redshift distributions, Fig.~\ref{fig_nz}. Figure~\ref{fig_xga} shows the results of power-law fits to the measured angular cross-correlations over our redshift range. The Limber inversion to a physical cross-correlation, $(x_0/r)^\gamma$, uses the redshift distributions of both sets of galaxies in the cross-correlation. The resulting $[x_0,\gamma]$ pairs are shown in Figure~\ref{fig_xg}. Comparison to Figure~\ref{fig_rg} shows that the $[x_0,\gamma]$ pairs are similar to the $[r_0,\gamma]$ pairs as a function of mass. The two lowest mass bins have switched positions with the cross-correlation of the lowest mass galaxies being stronger than the auto-correlation, although the difference is at about 2 standard deviation level in either $x_0$ or $x_0^\gamma$, which are the important quantities for pair counts. The difference arises from the best fit, since the point-by-point differences are all statistically equal for the two lowest mass correlations.

\section{LOW MASS GALAXY NUMBERS}

\begin{figure}
\epsscale{0.90}
\plotone{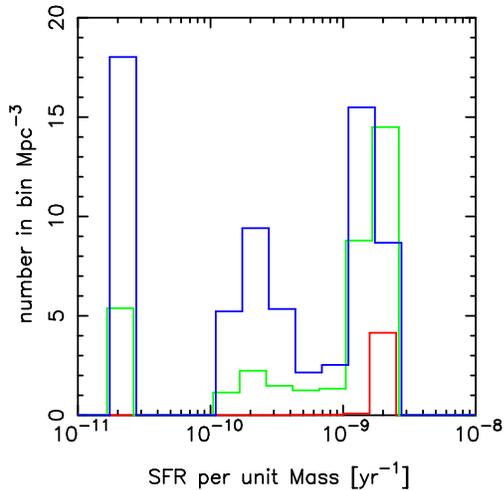}
\caption{The volume distribution of specific star formation rate for the $z=0.1-0.18$ (red), $z=0-0.05$ (green), and $z=0-0.03$ (blue) samples. The SFR is averaged over 0.5 Gyr and the method sets a limit of about $10^{-8}{\rm yr}^{-1}$ on the maximum value. 
\label{fig_nsS}}
\end{figure}

\begin{figure}
\epsscale{0.90}
\plotone{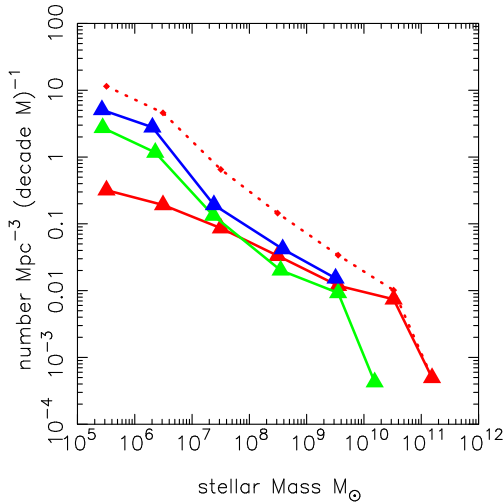}
\caption{The field volume density of galaxies as a function of mass for the $z=0.1-0.18$ (red), $z=0-0.05$ (green), and $z=0-0.03$ (blue) samples.
\label{fig_nM}}
\end{figure}

To measure the number of low mass satellite galaxies around a massive galaxy we need to determine the mean volume density of dwarfs. Our $z=0.1-0.18$ sample is entirely composed of low mass dwarfs with high star formation rates-—otherwise they would be too faint. Direct evidence that we are missing low star formation rates galaxies is shown in Figure~\ref{fig_nsS}, where we display the distribution of the specific SFR in the $10^5-10^6\msun$ bin, comparing the $z=0-0.03$, $0-0.05$ and $0.1-0.18$ redshift ranges and showing the increasing incompleteness. At redshift 0.02 our sample reaches $M_i\simeq -8$ mag and our 0.75\arcsec\ resolution corresponds to $200h^{-1}$ pc, sufficient to reach about 100 of the 120 dwarf galaxies in the local volume \citep{Belokurov:07, Gilmore:07, Sharina:08}. The simplest interpretation is that these three specific SFR (sSFR) distributions are consistent with a similar underlying sSFR distribution with increasing bias towards high specific star formation rates. There is a likelihood of some evolution in the sSFR distribution even over this small redshift range so it would be unwise to assume an invariant sSFR distribution to estimate the total numbers of low mass galaxies. On the other hand, these star formation rates are not so high as to lead to significant evolution of the overall numbers of galaxies in this mass range, so we use the density of low mass galaxies at low redshift as our best estimate. All three densities are shown in Figure~\ref{fig_nM}. If we shrink the bin to $z=0-0.02$ we find the densities at all masses rise, which suggests that the volume is being underestimated, about as expected given the median redshift error of 0.02.  The range we use $z=0-0.03$ does not significantly suffer from this over-estimation at higher mass so we believe that the densities are essentially correct. The steepening of the luminosity function at low mass has been remarked for the Virgo cluster and other nearby systems \citep{SBT:85, TH:02, Sabatini:03, RG:08}.

As a point of interest, a very simple approach to estimating the incompleteness is to use the specific SFR as an estimate of the duty cycle for visibility at high star formation rate. This will not work for high mass galaxies since the entire population is visible, independent of the star formation duty cycle. Normalizing to the highest mass bin, $M_0$, we estimate the numbers at lower mass as $n^\prime(M)=n(M)sSFR(M)/sSFR(M_0)$. The result is shown as the dashed line in Figure~\ref{fig_nM} and provides a reasonable estimate of the numbers, at about the factor of two level of precision. We do expect that a duty cycle model should be applicable at these low masses where the relatively short bursts of star formation cannot be sustained and the sSFR distribution appears to have a small, fairly narrow peak at the high sSFR, with most of the galaxies distributed over much lower sSFR. The analysis of Lee et al. (2008) finds a duty cycle of 6\% for a sample that is about a factor of ten more luminous than ours. Using the simple analysis above we find very similar duty cycles of 2.7\% for $10^5-10^6\msun$ and 4.1\% for $10^6-10^7\msun$. We note that the ratio of the star forming dwarfs to those which include the quiescent galaxies is 6.3\%, 6.8\%, 4.5\% in the three lowest mass bins, remarkably similar to Lee et al. (2008). Again this boosts our confidence that the redshifts are sufficiently accurate to place the galaxies into the appropriate bins.

%%The field volume density of galaxies as a function of mass for the $z=0.1-0.18$ (red), $z=0-0.05$ (green), $z=0-0.03$ (blue), $z=0-0.02$ (cyan), $z=0.1-0.35$ (magenta) and $z=0.35-0.5$ (yellow) samples. The effects of sample incompleteness are visible with increasing redshift. At $z<0.02$ redshift errors lead to an underestimate of the sample volume.

\section{FIELD GALAXY SATELLITE COUNTS}

\begin{figure}
\epsscale{0.90}
\plotone{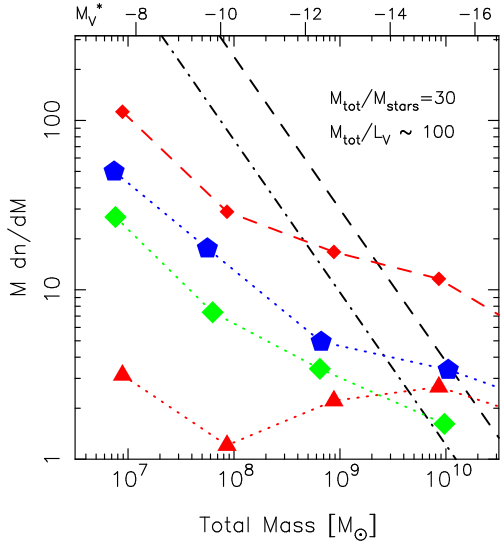}
\caption{The average number of galaxies as a function of mass inside a radius of $433$ kpc (H=73) for an adopted $M/M_\ast=30$, approximately $M/L_V\simeq 100$, for the $z=0.1-0.18$ (red), $z=0-0.05$ (green), and $z=0-0.03$ (blue) samples. We consider the $z=0-0.03$ sample to be largely complete to $M\simeq -8$, comparable to local group samples. An absolute luminosity scale for quiescent galaxies with $M_\ast/L_V=3$ is shown on the top axis.  The dashed lines are the Aquarius simulation results and Aquarius divided by 3.1 which approximate the Via Lactea result (dot-dashed line).
\label{fig_nhalo}}
\end{figure}

N-body simulations \citep{Aquarius, VL1, VL2} provide quantitative predictions for the numbers of low mass sub-halos around a Milky Way type galaxy. There is some disagreement over predicted numbers, which is likely largely due to the details of the input cosmological parameters, the density perturbation spectrum slope and normalization and the specific details of small numbers of realizations. Using our cross-correlation and density statistics we calculate $N(M,r) =4 \pi n(M) \int_0^r (x_0(M)/x)^\gamma x^2 dx $, where $r$ is set to the estimated 50 times critical density radius, 433\kpc\ (for h=0.73) which is the radius of comparison for the Aquarius and Via Lactea simulations. For comparison to the simulations we need to estimate the dark matter sub-halo mass associated with our dwarf galaxies. We use $M/M_\ast=30$, which for quiescent galaxies is $M/L_V\simeq 100$. The counts with this normalization are displayed in Figure~\ref{fig_nhalo}. Although very low luminosity dwarfs can have $M/L_V$ approaching 1000, these are too faint to enter into our sample \citep{SG:07}. Increasing the dark matter ratio a factor of three, to $M/M_\ast=100$, or a quiescent $M/L_V\simeq 300$, would alleviate the discrepancy of Fig.~\ref{fig_nhalo} slightly but not nearly enough to make any qualitative improvement. Similarly if we ascribe all our dwarfs with $M_\ast<10^7\msun$ to be resident within $10^7\msun$ halos \citep{Strigari08} the discrepancy is hardly changed.

The average number of satellite galaxies with $M_\ast>10^5\msun$ per massive galaxies is 78 and 39, for our field density estimates to $z=0.03$ and $0.05$, respectively. The systematic errors and field to field variance dominate the numbers so we will take our result to be $60\pm20$ satellites. To the same limit, which we take to be $M_V=-8$ mag, the Milky-Way has about 30 satellites. The duty cycle approach gives a total of about 175 satellites on the average.  Our assessment is that given the various uncertainties the agreement between our field estimates and the Milky Way counts is surprisingly good, with our estimates tending to be about a factor of two higher. The trend of an increasing discrepancy to lower mass is also seen. At a halo mass of $10^7\msun$ the numbers are a factor of 30 to 100 below the simulation results. The problem is one of the slope of the observed relation is much shallower, approximately $dN/d\ln{M}=-0.5$, than the predicted one $-0.9$, not a case of some mass scale at which the relationships diverge. The purely random statistical errors of about 20\% are about the size of the symbols.

\section{DISCUSSION}

This paper uses faint galaxy photometric redshifts to find very low stellar mass galaxies that are nearby. We estimate that at least 80\% and possibly 90\% of the redshifts are within the stated error of 0.02 in redshift, with the rest being fairly uniformly distributed over the entire redshift range searched, $z=0-3$. The analysis in this paper uses redshift bins that are larger than the errors and angular correlation analysis to minimize the impact of redshift errors. From The tests and comparisons in this paper establish the utility of the sample, but there is little doubt that obtaining spectroscopic redshifts and other follow-up studies would be of great interest. 

The correlation length for $10^5-10^6\msun$ galaxies is about 0.4\hmpc. There is no clear prediction for comparison to our measurements. Linear biasing theory calibrated at higher mass predicts a value about 76\% of the linear regime, about 3\hmpc, far above our measurement. The limited evidence from existing n-body simulations suggests that the nonlinear bias is likely to be much smaller than linear theory gives, and may even fall into the range of our measurements.

To estimate the number of satellite galaxies within the 50 times critical over-density radius Milky-Way, approximately 400 kpc, requires us to make a measurement of the field volume density of our dwarf galaxies. The cross-correlation sample at a mean redshift of about 0.15 is completely missing the low sSFR dwarfs, but they are visible in lower redshift samples which we therefore use to estimate the density. Combining the densities and the cross-correlations predicts the number of low mass galaxies, to $10^5\msun$ inside a large halo as approximately $60\pm20$, as compared to 30 for the Milky-Way to a comparable limiting stellar mass. The radial distribution of our dwarfs is much more centrally concentrated than the n-body simulation predictions which find that most of the objects are at large radius, whereas our cross-correlation, Fig.~\ref{fig_xcor}, is equivalent to an approximately linear rise with radius, and the slope of the relation does not vary with radius. Although the statistics do not allow a comparison of the radial slope, both the Milky-Way and M31 dwarfs are also much more centrally concentrated than the simulated sub-halos \citep{KGK,Willman04}. 

One of the possible explanations for the deficiency of dwarfs is that those low mass halos that did not collapse before reionization had their star formation largely suppressed forever \citep{Bullock:01}. The ``illuminated'' halos have numbers similar to the dwarf galaxy counts \citep{Strigari07}. The mass of dwarf galaxies is everywhere dominated by dark matter with the stars contributing relatively little to the binding energy. Stellar astrophysics adds energy to the gas which normally works to help drive gas out of the dwarf. The dwarf galaxy stellar populations are a mix of stars of every age, with significant ongoing bursts \citep{Mateo:98}. The evidence suggests that virtually all dwarfs cycle through bursting and passive stages (Lee et al. 2008). By extension, it would not be possible to stop low mass dark halos that did not happen to have stars formed near reionization to form stars at a later time. An observational argument is that the suppressed star formation would lead to a range of $M/L_V$ values that extended to infinity, whereas in our range of absolute magnitude more luminous than -8 mag the distribution appears to be well bounded.  Hence, the census of low mass dark halos is likely fairly complete within our mass limit.

It is possible that the total extent of the dark matter is much larger than the stellar contents of a dwarf galaxy \citep{MCS:05,PMN} in which case the mass will be underestimated. A dark matter mass to luminous mass ratio of approximately $1000$, roughly $M/L_V\simeq 3000$, at a stellar mass of $10^{5.5}\msun$ is required to agree with the Aquarius simulation, decreasing to dark to luminous mass ratio of $30$ at a stellar mass of $10^{8.5}\msun$. Such very large $M/L_V$ values have not been advocated in the observational literature, although the mass outside the stellar distribution could lead to such large values.  Our field density, about 5 Mpc$^{-3}$, of dark-to-stellar mass ratio 1000 halos imply an $\Omega_M \simeq [0.01,0.03,0.03,0.10]$ in the bins with mean mass $[10^{5.4},10^{6.3},10^{7.4},10^{8.6}]\msun$  which comes close to the total dark matter density but does not exceed it. 

The fact that the radial distribution of dwarfs are more centrally concentrated than the simulations imposes a dynamical limit on the $M/L_V$ of the satellite galaxies. Either the numbers need to rise by about 30 or the masses need to go up by about 30 times to bring the total counts inside the virial radius into basic agreement. However, both of these (if uncorrelated) will increase disk heating by a factor of 30,  leading to considerably more disk heating than the shallower distributions find \citep{Font:01,Ardi:03} likely at a level that would be unacceptable. 

Our overall conclusion is that we have found approximately twice as many dwarf galaxies per galaxy, $60\pm20$, as the 30 in the same stellar mass range as for the Milky-Way. For conventional $M/L_V\simeq 100$ (a dark-to-stellar mass ratio of 30) this then slightly alleviates the missing satellite problem, but still leaves the numbers about a factor of 30 short of the n-body predictions. Although there are no clear predictions for the two-point auto-correlation function in the non-linear regime for low mass galaxies, the very short correlation length of 0.4\hmpc\ is smaller than expected from the linear regime. The outcome of this paper is to extend the measurements of dwarf clustering to the field which results in a strengthening of the discrepancy with LCDM clustering expectations. 

Peebles (1989, 2007) has pointed out that dwarf galaxies, as a low bias population, should be a reasonable tracer of the dark matter and should therefore be present in the voids between massive galaxies, whereas it appears that the dwarf galaxies largely follow the massive galaxies. The fields appear remarkably uniformly populated with dwarf galaxies, Figure~\ref{fig_D1}, however it is important to recognize that we only measure clustering to approximately 0.4\hmpc, not the 10\hmpc\ scale and larger that characterize voids.

\acknowledgements

This paper is based on observations obtained with MegaPrime/MegaCam, a joint project of CFHT and CEA/DAPNIA, at the Canada-France-Hawaii Telescope (CFHT) which is operated by the National Research Council (NRC) of Canada, the Institut National des Sciences de l'Univers of the Centre National de la Recherche Scientifique (CNRS) of France, and the University of Hawaii. This work is based in part on data products produced at the Canadian Astronomy Data Centre as part of the CFHT Legacy Survey, a collaborative project of NRC and CNRS.  Canadian collaboration members acknowledge support from NSERC and CIFAR.

{\it Facilities:} \facility{CFHT}.

\end{document}